\documentclass{aa}
\usepackage{amsmath,graphicx,amssymb}
\usepackage{txfonts}
\usepackage{natbib}
\bibpunct{(}{)}{;}{a}{}{,}
\def\urltilda{\kern -.15em\lower .7ex\hbox{\~{}}\kern .04em}
\newcommand{\oc}{\textit{O-C}}
\newcommand{\w}{\citet{wilson14}}
\newcommand{\hvezda}{XY\,Boo}
\newcommand{\lef}{\lambda_{\mathrm{eff}}}
\newcommand{\zav}[1]{\left(#1\right)}
\newcommand{\hzav}[1]{\left[#1\right]}

\newlength\staretab
\makeatletter
\def\sgn{\mathop{\operator@font sgn}\nolimits}
\makeatother

\begin{document}

\title{Phenomenological modelling of eclipsing system light curves}
\titlerunning{Phenomenological modelling of eclipsing system light curves}

\author{Zden\v{e}k Mikul\'a\v sek}
\authorrunning{Z.~Mikul\'a\v sek}
\offprints{Zden\v ek~Mikul\'a\v sek,\\
\email{mikulas@physics.muni.cz}}

\institute{Department of Theoretical Physics and Astrophysics,
           Masaryk University, Kotl\'a\v{r}sk\'a 2, CZ\,611\,37, Brno, Czech Republic}

\date{Received / Accepted }

\abstract{The observed light curves of most eclipsing binaries and stars with transiting planets can be well described and interpreted by current advanced physical models which also allow for the determination of many physical parameters of eclipsing systems. However, for several common practical tasks there is no need to know the detailed physics of a variable star, but only the shapes of their light curves or other phase curves.}{We present a set of phenomenological models for the light curves of eclipsing systems.}
{We express the observed light curves of eclipsing binaries and stars, transited by their exoplanets orbiting in circular trajectories, by a sum of special, analytical, few-parameter functions that enable fitting their light curves with an accuracy of better than 1\%. The proposed set of phenomenological models of eclipsing variable light curves were then tested on several real systems. For XY Bootis, we also compare in details the results obtained using our phenomenological modelling  with those found using available physical models.}{We demonstrate that the proposed phenomenological models of transiting exoplanet and eclipsing binary light curves applied to ground-based photometric observations yields results compatible with those obtained by the application of more complex physical models.}{The suggested phenomenological modelling appears useful to solve a number of common tasks in the field of eclipsing variable research.}

\keywords {stars: variables -- stars: eclipsing:  -- exoplanets}

\maketitle

\section{Introduction}

There are two groups of astrophysical tasks standardly solved by the analysis of observations of eclipsing binaries (EBs) and stars with transiting extrasolar planets. We can derive the physics of the present state of a system and its components -– namely the dimensions and geometry of the system, outer characteristics of both eclipsing bodies such as their radii, shapes, masses, temperatures, limb darkening, gravity darkening/brightening, gravitational lensing, albedos, spottiness, pulsation, and parameters of possible streams and disks. Required information is extracted by means of various, more or less sophisticated physical models of double systems \citep[e.g.][]{wilson,kalmil,had,BM3,prsa05,prsa11,prib12} applied to numerous and precise data of all kinds obtained by contemporary observational methods and approaches developed for this purpose.

Equally important are the description and classification of light curves (LCs), or studies of the evolution of the systems, within the time scale of decades. These usually very tiny changes in light variation parameters (typically the period) may provide information on e.g. the rate of mass exchange between interacting components \citep[e.g.][]{zhudd,zhubs,miktatry}, the presence and characteristics of possible invisible bodies \citep[stars, planets, e.g.][]{van,qian3} in the system, or the degree of the mass concentration in stellar interiors (internal structure constants) if we study apsidal motion in some eccentric EBs \citep[e.g.][]{kopal,claret,zasche13}, etc.

In period analyses of EBs we do not need all the information about the physics of the system, but only good templates of phase curves of all data we want to analyse \citep{miktatry}. Such template phase curves (typically light curves in various photometric bands) can be obtained in principle by means of the sophisticated versions of physical EB models applied to the complete set of observational data \citep[see in][]{van,wilson14} or to their best parts \citep[e.g.][]{zasche14,zasche15}.  There is also the possibility of using the best observed LCs themselves \citep{pribulla}.
We offer another alternative using relatively simple phenomenological (mathematical) modelling of observed data variations.

Our aim is to establish a general model of light curves of \textit{eclipsing systems -- ES} (both eclipsing binaries and stars with transiting planets) that could fit the LCs with an accuracy of 1\% of their amplitudes or better that can be applied to the majority of observed eclipsing systems \citep[the concept of the model is outlined in][]{mikcon}. Such a model could be used for most LC description tasks and detailed period analysis, for which other sources of phase information could be also used, especially individual eclipse timings and radial-velocity (RV) curves.

The remainder of this paper is organised as follows: in Sect.\,\ref{gen} we describe the general properties of phase variation of periodically variable objects, Sect.\,\ref{zero} specifies general properties of eclipsing system LC models, and Sect.\,\ref{fenec} presents the phenomenological model of eclipsing system light curves. Sect.\,\ref{solution} is devoted to model parameters and searching for them during analysis, in Sect.\,\ref{xyperchng} we compare the results of the modelling XY\,Boo observations processed by means phenomenological and physical \citep{wilson14} methods, and in Sect.\,\ref{conclusions} we summarise and discuss our results of the phenomenological modelling of the eclipsing systems.

\section{Periodically variable objects}\label{gen}

Eclipsing binaries and stars with transiting exoplanets periodically change their light mainly due to regular mutual eclipses of components (transits and occultations) and proximity effects (tidally induced ellipticity of components and reflection). The period of these ES variations corresponds to the observed orbital velocity of the system; the shape of light curves, dominated by eclipses and proximity effects, is hence more or less constant (for details see Sect. \ref{zero}). That is why ESs are ranked into a group of periodically variable stars, or objects in general.

Most of the variations of periodic variables are more or less cyclic with an observed \textit{instantaneous period} $P(t)$ which is usually strictly constant or slightly variable. The period itself and its development over time can not be observed directly, but both can be derived through the analysis of time series of light changes or extremum timings. For that purposes it was useful to introduce a monotonically rising  \textit{phase function} $\vartheta(t)$, as a sum of the epoch $E(t)$ and the phase in the common usage $\varphi$ and its inversion function $t(\vartheta)$ \citep{mik901}.

Functions $\vartheta(t)$ are tied with the instantaneous observed period $P(t),\,P(\vartheta)$ by the following differential equations \citep[see][]{kalimero,mik901,mikgra}:
\begin{equation}\label{phasefundef}
\frac{\mathrm{d}\vartheta(t)}{\mathrm{d}t}=\frac{1}{P(t)},\ \vartheta(M_0)=0;\quad \frac{\mathrm{d}t(\vartheta)}{\mathrm{d}\vartheta}=P(\vartheta);\ t(0)=M_0,
\end{equation}
where $M_0$ is the origin of counting of epochs (for ESs the time of the basic primary minimum).

Using $t(\vartheta)$ we can predict the zeroth phase time $\mathit{\Theta(E)}$ for the epoch $E$ (primary minimum timing according to the ephemeris Eq.\,(\ref{phasefundef})), $\mathit{\Theta(E)}=\mathit{\Theta}(\vartheta=E)$.

\subsection{Period models. Phase and time shifts.} \label{phase}

The linear period model supposes the period $P(t)$ of the variable object to be constant. The corresponding linear phase function $\vartheta_1(t,M_0,P_0)$ and its inversion $t_1(\vartheta,M_0,P_0)$ are then given by:
\begin{equation}\label{linmod}
 \vartheta_1(t)=(t-M_0)/P_0,\quad t_1(\vartheta)=M_0+P_0\,\vartheta.
\end{equation}
A possible tiny modulation $\Delta P(t)$ of the basic period $P_0$ causes a detectable phase function shift $\Delta \vartheta(t)$ in light curves and shifts $\Delta t(\vartheta)=\Delta \mathit{\Theta}(E)$ in LC extrema timing.
\begin{equation}
P(t)=P_0+ \Delta P(t);\ \ \vartheta(t)=\vartheta_1+\Delta \varphi (t);\ \
t(\vartheta)=t_1+\Delta t(\vartheta).\label{shifts}
\end{equation}
Combining definitions in Eq.\,\ref{shifts} with the fact that functions $\vartheta(t)$ and $t(\vartheta)$ are mutually inverse, we obtain the following relations
\begin{align}
&\Delta t = -P_0\,\Delta\vartheta; \quad \Delta P(\vartheta)=\frac{\mathrm d \Delta t(\vartheta)}{\mathrm d \vartheta}=\frac{\mathrm d \hzav{t(\vartheta)-t_1(\vartheta)}}{\mathrm d \vartheta};\label{relat}\\
&\Delta \vartheta(t)\doteq\int_{M_0}^t \hzav{\frac{\Delta P(\tau)}{P_0^2}
-\frac{\zav{\Delta P(\tau)}^2}{P_0^3}}\,\mathrm{d}\tau \doteq \int_{M_0}^t \frac{\Delta P(\tau)}{P_0^2}\,\mathrm{d}\tau. \label{deltafi}
\end{align}
The last approximation in the Eq.\,(\ref{deltafi}) is valid for all known EBs, including SV~Cen with a record-breaking decrease of its orbital period by $\dot{P}/P_0=-2.36(5)\times10^{-5}$\,yr$^{-1}$ \citep[][]{svicen}.

The time development of phase function shifts $\Delta \vartheta(t)$ can be derived by the analysis of light curves, whilst $\Delta \mathit {\Theta}(E)$, can be revealed using standard \oc\ diagrams constructed by means of extrema timings.

\subsection{Basic period models. Finding of \oc\ shifts}\label{permods}

If the time development of the period $P(t)$ is continuous and smooth we can express it in the form of the Taylor polynomial with the centre at $t=M_0$, $\vartheta_1=(t-M_0)/P_0$
\begin{equation}
\textstyle P(t)=P_0+P_0\dot{P}_0\,\vartheta_1 +P_0^2\ddot{P_0}\frac{\vartheta_1^2}{2!}+\ldots+ P_0^k \frac{\mathrm d^k\!P_0}{\mathrm dt^k}\frac{\vartheta_1^k}{k!}\ldots \label{perlaurin}
\end{equation}
Using Eqs. (\ref{linmod}), (\ref{shifts}), (\ref{relat}), and simplified version of Eq. (\ref{deltafi}) we obtain:
\begin{align}
\vartheta&\textstyle\simeq\vartheta_1 -\dot{P_0}\,\frac{\vartheta_1^2}{2!}-
P_0\ddot{P}_0\,\frac{\vartheta_1^3}{3!}-\ldots - P_0^{k-1} \frac{\mathrm d^k\!P_0}{\mathrm dt^k}\frac{\vartheta_1^{k+1}}{(k+1)!}\ldots \label{thetalaurin}\\
t&\textstyle\simeq M_0\!+\!P_0\,\vartheta+P_0\dot{P}_0\frac{\vartheta^2}{2!}\!+\! P_0^2\ddot{P}_0\frac{\vartheta^3}{3!}\!+\!\ldots\!+\!P_0^k \frac{\mathrm d^k\!P_0}{\mathrm dt^k}\frac{\vartheta^{k+1}}{(k+1)!}\ldots \nonumber     \\
\mathit\Theta&\textstyle\simeq M_0\!+\!P_0\,E+P_0\dot{P}_0\frac{E^2}{2!}\!+\! P_0^2\ddot{P}_0\frac{E^3}{3!}\!+\!\ldots\!+\!P_0^k \frac{\mathrm d^k\!P_0}{\mathrm dt^k}\frac{E^{k+1}}{(k+1)!}\ldots \label{Thetalaurin}
\end{align}
Similarly we can establish other arbitrarily complex period models of $\vartheta(t)$ determined by a set of free parameters including also cyclic period modulation of the  light time effect (LiTE) caused by another body in the system \citep{mikar,liska} or the apsidal motion.

The real shape of the phase curve can also be approximated using so-called \oc\ time shifts of the observed phase curves versus the predicted light curve (LC) derived by the period model with fixed parameters (typically $P_0,M_0$) expressed in time units (usually in days). Let us divide the whole time interval covered by observations into $n_{\mathrm{OC}}$ appropriate time intervals (typically nights or seasons). The phase function $\vartheta(t,r)$ during a certain $r$-th time interval with the $(\oc)_r$ time shift is then given by the formula \citep{miklijiang}
\begin{equation}\label{oc}
\vartheta(t,r,\vartheta_{\mathrm c},\{\oc\})=\vartheta_{\mathrm c}(t) - \sum_{s=1}^{n_{\mathrm{OC}}} \,\eta_{rs}\,\frac{(\oc)_s}{P},
\end{equation}
where $\vartheta_{\mathrm c}(t_i)$ is a predicted phase function calculated by an appropriate period model at the time $t_i$, $\{\oc\}$ is a set of all $n_{\mathrm{OC}}$ values of the found $(\oc)_r$ time shifts versus this model. The symbol $\eta_{rs}$ represents a discrete function (a table or a matrix), which for each individual $i$-th observation from our data assigns either 1, if its order number of the interval $r_i=s$,  or 0, if $r_i\neq s$.

The dependence of \oc\ values on the epoch serves as a common \oc\ diagram, the basic tool for the period analysis. Using the found individual \oc\ values and their uncertainties we can calculate a set of so-called virtual minima timings \citep[][or Sect. \ref{xyuniefem} in this paper]{miklijiang,mikcu}. Virtual minima timings can be combined with others derived e.g. by other techniques \citep[e.g.][]{brat,mandel,kwee,mikext,mikKW,zasche14,zasche15}.

\section{General properties of an eclipsing system light curve model}\label{zero}

\subsection{Instrumental term of an observed light curve}\label{instrum}

The observed light curve (or its segment) of a chosen eclipsing system in a particular colour of an effective wavelength $\lef$ is defined by a time series $\{t_{ri},y_{ri}\}$ obtained during an observational interval $r$ (observing night, part of it or season). The $r$-th subset of observational data can be generally modelled by the function $Y_r(t,\lef)$, expressed in magnitudes\footnote{The presented models may also be applied for expressing light intensity variations, but the treatment of data in the magnitude domain is more straightforward and the accuracies of results are almost the same.}:
\begin{equation}
Y_r(t,\lef)=Y_{0r}(t,\lef)+F(\vartheta,\lef); \label{general}
\end{equation}
where $Y_{0r}(t,\lef)$ is an additive term removing instrumental trends and concurring the observed magnitudes or magnitude differences $\{y_i\}$, whilst
$F(\vartheta,\lef)$ is an intrinsic light curve model function of the phase function $\vartheta$ (see Sect.\,\ref{phase}) and the effective wavelength $\lef$, free of instrumental and observational shifts and trends. The function $F(\vartheta,\lef)$ is normalized so that its mean value without eclipses is equal to zero. The function $Y_{0r}(t,\lef)$ can be approximated by a linear combination of $g_r$ dimensionless functions of time $t$, $\mathit{\Xi}_j\,(t)$, with magnitude-like coefficients $m_{0rj}(\lef)$:
\begin{equation}
Y_{0r}(t,\lef)\simeq \sum_{j=0}^{g_r} \,m_{0rj}\,(\lef)\ \mathit{\Xi}_{j}\,(t),\label{instrument}
\end{equation}
where $\mathit{\Xi}_j\,(t)$ can be e.g. normalised polynomials $\mathit{\Xi}_{rj}\,(t)=[(t-\bar{t_r})/\sigma(t_r)]^j$ ($\sigma(t_r)$ is the weighted variance of observational times in the segment $r$) or Legendre polynomials or special quasi orthogonal functions combining polynomials with harmonic functions \citep[see in][]{mikgr}. The set of coefficients $\{m_{0rj}\,(\lef)\}$ is found together with parameters describing the model function $F(\vartheta,\lef)$.

\subsection{Bases and limitations of the phenomenological model of eclipsing system light curves}\label{bases}

Light curves of ESs are nearly periodic functions and it would be natural to express them in the form of Fourier series \citep[see e.g.][]{ruc,kalmil,selam,nedor,and12}. This concept proves its worth in many types of extrinsic periodically variable stars, especially in the case of rotating variables with photometric spots on their surfaces \citep{north,mikzoo} or non-eclipsing (e.g. purely elliptical) double stars, where we manage with harmonic polynomials of a low degree \citep[e.g.][]{kalmil}.

However, it is generally known that the presence of eclipses in LCs asks for the use of harmonic polynomials of a rather high degree if a proper fit of the observed LC is required. Their usage is badly influenced by departures from the ideal equidistant distribution of observations according to the orbital phase. Even small phase gaps are then filled with unreal LC artifacts.

\begin{figure}
\centering \resizebox{0.94\hsize}{!}{\includegraphics{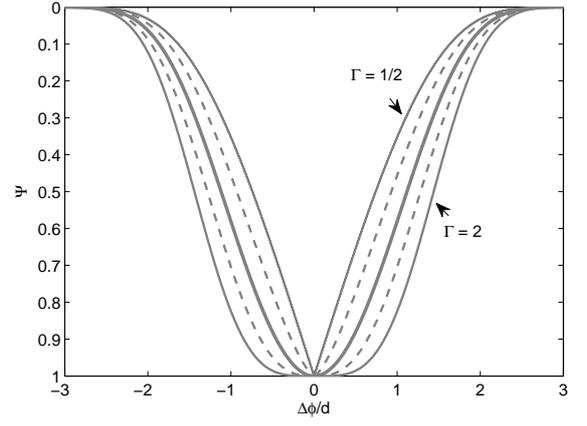}}
\caption{\small{Examples of elementary model LC $\mathit{\Psi}=F_{\mathrm e}/A_k$, for $C=0$, see Eq.\,(\ref{modelecl}) allowed by our phenomenological EB models. Parameter $d$ describes the duration of the eclipse, $\mathit{\Gamma}$ is a parameter expressing the kurtosis of the LC.}} \label{cecka}
\end{figure}

It seems that it is better to use properly selected phenomenological models of LCs of eclipsing systems described by a few parameters. Several more or less successful attempts on how to model these LCs have been proposed e.g. by \citet{ces71,chol}, see also the reviews and references in \citet{and12,chra}, Table\,\ref{srov}, and Fig.\,\ref{cecka},\,\ref{arsrov}.

Theory is able to explain the observed periodic light variations of an ES in general and in detail as the result of alternating mutual eclipses of components of the system, non-isotropic radiation of orbiting components, caused by their close proximity, and by unevenly distributed photometric spots on rotating components \citep[e.g.][]{BM3,hilda,pribulla}. The periodicity of light changes caused by mutual eclipses and proximity effects, as well as variations connected with the rotation of synchronously rotating spotted components is dictated by their observed orbital motion. These changes are periodic with an instantaneous period\footnote{Observed orbital period may change due to possible transfer of matter between components or light time effect.} $P(t)$, the shapes of light curves remain more or less constant for decades.

For purely geometrical reasons the prevailing majority of EBs ranks among relatively close, and hence tidally interacting systems, where the processes of synchronizing of the components'rotation and orbit circularization are strong and effective \citep[][and references therein]{zahn,goldma}. That is why the rotations of components are usually synchronous and more than 80\,\% of their orbits are pretty circular \citep[see CALEB,][]{BM3}. In the further introductory text we will concentrate mainly on systems with more or less circular orbits\footnote{Applying so-called \textit{phase rectification} (Mikul\'a\v{s}ek et al. in preparation) we are also able to solve eccentric systems. The technique symmetrizes light curves of eclipsing binaries with components moving unevenly on their eccentric orbits by the rectification of their phase functions. The method of phase rectification enables also the effective analysis of apsidal EB motion.} \label{rector} and constant light curves\footnote{This assumption is fulfilled only partially. We can mention cyclical variations of instantaneous LCs with other than orbital period as the gradual change of the ES geometry due to asymptotic motion of double systems orbiting on eccentric orbits, possible asynchronous rotation of spotted components in wide systems, and possible pulsations of the components. Eclipsing binary light curves may also vary erratically because of chromospheric activity (see e.g. Sect.\,\ref{xyphasecrv}), time-dependent spottiness of the components, or changes in streams or disks around the stars.
Neglecting of above mentioned effects introduce as a rule some extra noise in the period analyses and deteriorate the accuracy of the determination model parameters.}.

The intrinsic one-colour light curve function (expressed in magnitudes) of an eclipsing system is a periodic function $F(\vartheta,\lef)$ that can be approximated as the sum of three more or less independent terms \citep[see also in][]{and12}:
\begin{equation}\label{EBbas}
F(\vartheta,\lef)=F_{\mathrm e}(\vartheta,\lef)+F_{\mathrm p}(\vartheta,\lef)+F_{\mathrm c}(\vartheta,\lef),
\end{equation}
where $F_{\mathrm{e}}(\vartheta,\lef)$ describes a light curve of eclipses  ($F_{\mathrm{e}}\equiv 0$ outside of eclipses), whilst $F_{\mathrm{p}}(\vartheta,\lef)$ and $F_{\mathrm{c}}(\vartheta,\lef)$ express contributions of proximity and \citet{con} effects without eclipses ($\overline{F_{\mathrm p}}=\overline{F_{\mathrm c}}=0$). The mathematical models are formulated and discussed in Sect.\,\ref{proxicon}.

\section{Phenomenological model of eclipsing system light curves }\label{fenec}

\subsection{Model of one-colour light curves}\label{monmodel}

\subsubsection{Eclipses}\label{monec}

The essential feature of all ES light curves are two nearly symmetrical depressions caused by mutual eclipses of synchronously rotating stellar or planetary components. The profiles of both minima are complex functions determined primarily by the geometry of the system and the relative brightness of components in a given spectral region centered at the effective wavelength $\lef$. The contribution of eclipses $F_{\mathrm{e}}(\vartheta,\lef)$  to an ES light curve can be approximated by a sum of two special periodic functions of phase function $\vartheta$. In the case of circular orbits eclipses are exactly symmetrical around their centres at phases $\varphi_{01}$ and $\varphi_{02}$. If we put the origin of the phase function $M_0$ at the time of the primary minimum, then $\varphi_{01}=0,\ \varphi_{02}=0.5$.

The model function was selected so that it describes as aptly as possible those parts of LCs that are in the vicinity of their inflex points, where their slopes are maximal. The functions are parameterised by their widths $D_1,\,D_2$, eclipse LC kurtosis coefficients $\mathit{\Gamma}_1,\,\mathit{\Gamma}_2$, dimensionless correcting factors $C_1, C_2$, and central depths $A_1(\mathit{\lef}),\, A_2(\mathit{\lef})$:
\begin{align}\label{modelecl}
&\displaystyle F_{\!\mathrm{e}}(\vartheta,\!\lef)=\! \sum_{k=1}^{n_{\mathrm{e}}}\!A_{k}\zav{1\!+\!C_{\!k}\!\frac{\varphi_k^2}{D_k^2}} \left\{1\!-\!\left\{1\!-\!\exp\!\hzav{1\!-\!\cosh\!\zav{\frac{\varphi_k}{D_k}}} \right\}^{\mathit{\Gamma\!_k}}\right\},\nonumber\\
&\varphi_k=\vartheta-0.5\,(k-1) -\mathrm{round}\hzav{\vartheta-0.5\,(k-1)},
\end{align}
where the summation is over the number of eclipses during one cycle, $n_{\mathrm e}$: $n_{\mathrm e}=2$ or $n_{\mathrm e}=1$ (the common situation for exoplanet transits). Each eclipse in a given colour is thus described by only four parameters - its depth $A$, width $D$, kurtosis $\mathit{\Gamma}$, and the correcting parameter $C$.

\begin{figure}
\centering \resizebox{1\hsize}{!}{\includegraphics{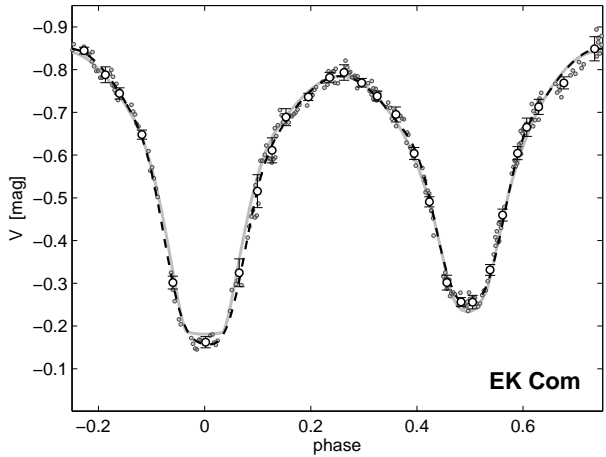}}
\caption{\small{The fit of the $V$ curve of the overcontact spotted EB EK\,Com (the secondary minimum is a transit). The LC of the close binary is affected by O'Connell and proximity effects. The synthetic model curve is depicted by a solid line, the phenomenological LC fit is depicted by the dashed line. For more information see Sect.\,\ref{select} and Table\,\ref{prm}.}}\label{ekcom}
\end{figure}

In the case of eclipsing binaries with two minima in a cycle ($n_{\mathrm e}\!=\!2$) we need eight parameters, but sometimes the number of needed parameters can be smaller. Inspecting the parameters $D,\ \mathit{\Gamma}$, and $C$ for both eclipses of many EBs we have concluded that they are as a rule nearly the same: especially $D_1\!\cong\!D_2,\ \mathit{\Gamma}_1\!\cong\!\mathit{\Gamma}_2$, and $C_1\!\cong\! C_2$. So we usually need only five monochromatic parameters ($A_1,A_2,D,\mathit{\Gamma},C$). The parameter $C$ is mostly comparable with its uncertainty, so we can neglect it entirely. Then we need just four parameters! On the other hand, in EBs with totalities we see that the bottoms of their occultations are flat whilst transits are convex. It can be described by introducing of different parameters $C_1, C_2$ (see the case of EK~Com in Table\,\ref{prm}, Fig.\,\ref{ekcom}).

\begin{figure}
\centering \resizebox{0.86\hsize}{!}{\includegraphics{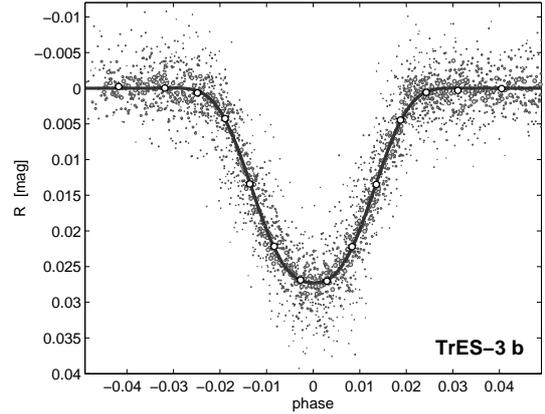}}
\caption{\small{The simultaneous fit to 15 transits of extrasolar planet TrES-3b corrected for trends (see Eq.\,(\ref{general})). The fit parameters according to Eq.\,(\ref{modeltrs}): $A=0.02725(19)$\,mag, $D=0.0117(7)$, $C=-0.17(7)$, and $\mathit{\Gamma}=1.58(12)$, the parameter $K$ was not introduced. The black line is the fit, circles are the normal points, grey circles indicate individual measurements with the area proportional to their weights.}} \label{tres3}
\end{figure}

The LCs of the exoplanet transits ($n_{\mathrm e}=1$) need only four parameters ($A,D,\mathit{\Gamma},C$ - see Fig.\,\ref{tres3}), in cases of very precise measurements we add another dimensionless parameter $K$ \citep{mikcon}:
\begin{equation}\label{modeltrs}
\displaystyle F_{\!\mathrm{e}}=A\zav{1\!+\!C\frac{\varphi^2}{D^2}\!+ \!K\frac{\varphi^4}{D^4}} \left\{1\!-\!\left\{1\!-\!\exp\hzav{1\!-\!\cosh\zav{\frac{\varphi}{D}}} \right\}^{\mathit{\Gamma}}\right\}.
\end{equation}
Testing several dozens of LCs of various types of eclipsing systems we found that the standard deviation of the fit is typically well bellow one per cent. The only minor inconvenience is the existence of a spike (a jump in derivatives) in the mid-eclipses for LCs with $\mathit{\Gamma}<1$ (see Fig.\,\ref{cecka}).

\subsubsection{Proximity effects. O'Connell effect}\label{proxicon}

Light variations of EBs caused by eclipses are usually modified by asphericity of the components, effects of gravity  darkening/brightening and mutual irradiation. All these \emph{proximity effects} are the manifestation of the interaction between the components acting, namely, in close systems. Contrary to eclipses the proximity effects modify LCs permanently, in each phase.

Light curves of some eclipsing binaries are influenced by the O'Connell effect that results in the asymmetry of some LCs of close EBs, manifesting as the difference in light maxima between eclipses. The standard explanation for this is the presence of one or more cool or hot spots on the surface of one of the synchronously rotating components or by asymmetrically distributed circumstellar material in the system \citep[e.g.][and citations therein]{wilsey,pribulla}. The amount and the sign of the O'Connell effect vary with time \citep{beaky}. The effect is also wavelength dependent -- in blue it is usually stronger, but it is not a rule \citep{pribulla}.

The contribution of proximity effects $F\!_{\rm p}(\vartheta)$ should be an even function symmetric with the phases 0.0 and 0.5 and consequently they can be satisfactorily well expressed as a linear combination of $n_{\rm p}$ elementary cosine functions $\cos(2\,\pi\,\vartheta),\,\cos(4\,\pi\,\vartheta),\,\cos(6\,\pi\,\vartheta),\ldots $ . The even terms are the consequence of the ellipticity of tidally interacting components, whilst the odd terms result from the differences between the near and far sides of components. As a rule we can limit ourselves only to the first two or three terms in the $F\!_{\rm{p}}$ \citep{rusmer,kalmil}.  The O'Connell effect contribution $F_c(\vartheta)$  can be well modelled by a simple sinusoid \citep{davmil,wilsey}:
\begin{equation}\label{proxcon}
 F\!_{\rm p} =\!\sum_{k=n_{\mathrm e}+1}^{n_{\mathrm p}+
n_{\mathrm e}}\! A_k\cos\hzav{2\pi(k\!-n_{\mathrm e})\vartheta},\
  F\!_{\rm c}=\!\sum_{k=n_{\mathrm p}+n_{\mathrm e}+1}^{n_{\mathrm c}+
n_{\mathrm p}+n_{\mathrm e}}\! A_k\sin(2\pi\vartheta),
\end{equation}
where $n_{\mathrm p}$ is the number of terms in $F\!_{\rm p}(\vartheta)$: $n_{\mathrm p}=0,1,2,3,\ldots$, $n_{\mathrm c}=0$, if the O'Connell asymmetry is not present\footnote{If $p>q$ then $\sum_{k=p}^q\,h_k=0.$}, else $n_{\mathrm c}=1$.

\begin{table}\small
\begin{center}
\caption{\small{Parameters describing LCs of several eclipsing binaries. }}\label{prm}
\begin{tabular}{ccccc}
\hline
  Name &AR\,Aur & EK Com & AV Del & 477 Lyr\\
   \hline
  $g_{\mathrm{a}}$ & 9 & 10 & 13 & 8 \\
  filter & $B/V$& $V$ & $BVRI$ & $V$ \\
  $n$ &686/687&247&545&294\\
  $D$ & 0.0168(3)& 0.052(3)& 0.0396(7) &0.0219(5) \\
  $\mathit{\Gamma}$ &0.619(10)& 1.7(4)&0.82(5)&0.79(4)\\
  $a_{11}$ &0.691/0.681&0.331(12)&1.153(11)&1.557(14)\\
  $a_{12}$ &--&--&1.17(4)&--\\
  $a_{21}$ &0.551/0.564&0.289(9)&0.123(3)&0.078(5)\\
  $a_{22}$ &--&--&-0.188(13)&--\\
  $a_{31}$ &-0.0022(5)&0.023(5)&--&0.321(2) \\
  $a_{32}$ &-0.022(4)&--&--&--\\
  $a_{41}$ &0.0028(9)&0.150(5)&0.050(2)&-0.035(2)\\
  $a_{42}$ &--&--&-0.044(1)&--\\
  $a_{51}$ &--&--&--&-0.005(2)\\
  $a_{61}$ &--&0.032(2)&--&--\\
  $C$ &-0.25(2)&--&-0.24(7)&-0.342(20)\\
  $C_{1}$ &--&0.10(17)&--&--  \\
  $C_{2}$ &--&-0.14(9)&--&--  \\
  $\rho$ & 1.05& 0.73& 1.05&0.31  \\
  \hline
  type&OC&SD&D&D\\
  $P$&$4\fd13169$&$0\fd26669$&$3\fd8534$&$0\fd47173$\\
  $r_1$ &0.098& 0.524& 0.198&0.077  \\
  $r_2$ &0.100& 0.318& 0.343&0.219  \\
  $T_{\mathrm{eff}1}$ &11\,100\,K& 5000\,K& 6000\,K&60\,000\,K \\
  $T_{\mathrm{eff}2}$ &10\,600\,K& 5300\,K& 4275\,K&6500\,K \\
  $i$ &$88.5^{\circ}$& $88.5^{\circ}$& $81.3^{\circ}$&$65.8^{\circ}$\\
  Fig. &\ref{arprim}& \ref{ekcom}&\ref{avdel}&\ref{477lyr}\\
  \hline
\end{tabular}
\end{center}
\tiny{$g_{\mathrm{a}}$ is the number of the parameters used for the description of a LC, $\rho=\delta_{\mathrm{phen}}/\delta_{\mathrm{phys}}$, where $\delta_{\mathrm{phen}}$ and $\delta_{\mathrm{phys}}$ are the uncertainties in the determination of the zero phase time according to phenomenological and physical models. The used `hi-fi' LC model: $F(\vartheta,\mathit{\Lambda})=\sum_{k=1}^2\sum_{j=1}^3\,a_{k\!j}\  \mathit{\Lambda}^{j-1}\
[1+C_k\,(\varphi_k/D)^2]\ \{1-\{1-\exp\hzav{1-\cosh\zav{\varphi_k/D}} \}^{\mathit{\Gamma}}\}$ $+\sum_{k=3}^5\sum_{j=1}^2\,a_{k\!j}\,\mathit{\Lambda}^{j-1} \cos\hzav{2\,\pi\,(k\!-\!2)\,\vartheta}+a_{60}\,\sin(2\,\pi\,\vartheta).$ EB types: OC - overcontact, C - contact, SD - semidetached, D - detached, $P$ is the period, $r_1,r_2$ are the relative radii of the components, $T_{\mathrm{eff}1},T_{\mathrm{eff}2}$, their effective temperature, $i$ is the orbit inclination. The parameters were taken from authors cited in Sect.\,\ref{select}.}
\end{table}

\begin{figure}
\centering \resizebox{0.94\hsize}{!}{\includegraphics{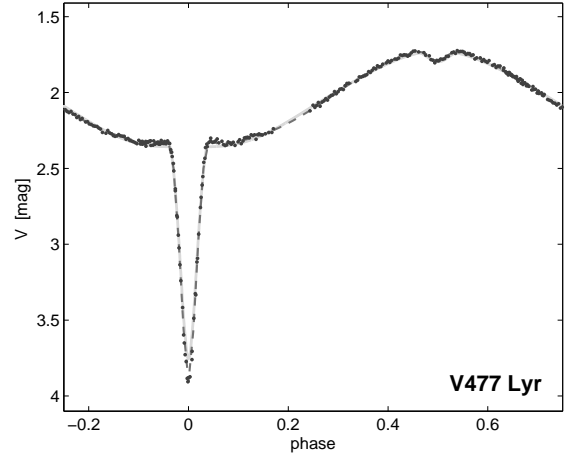}}
\caption{\small{$V$ light curve of a famous EB consisting of a very hot nucleus of a planetary nebula and solar type star. For the fit of the LC strongly affected by proximity effects we need only 8 parameters phenomenological. For details see Sect.\,\ref{select} and Table\,\ref{prm}. }} \label{477lyr}
\end{figure}

\begin{figure}
\centering \resizebox{0.85\hsize}{!}{\includegraphics{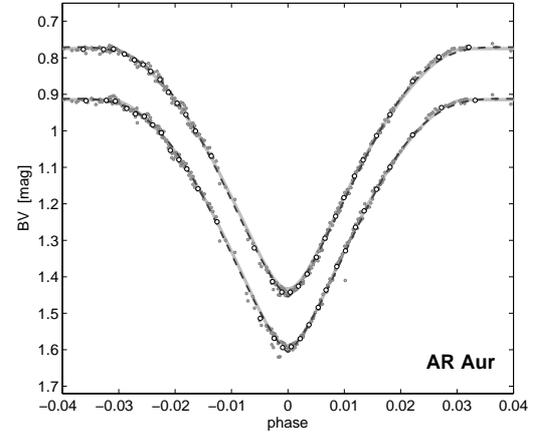}}
\caption{\small{The fit of the primary minimum of the detached EB AR\,Aur in $V$ and $B$, represented by normal points (circles), by physical (grey solid lines) and phenomenological (dashed lines) LCs (see\,Table\,\ref{prm}).}} \label{arprim}
\end{figure}

The $V$ light curve of the close eclipsing binary EK Com (see Fig.\,\ref{ekcom}) with apparent O'Connell effect is described by nine parameters: $A_{1,2,3,4,5},D,C_{1,2},\mathit{\Gamma}$,\, $n_{\rm e}=2$,\,$n_{\mathrm p}=2,\,n_{\mathrm c}=1$.
The uncommon $V$ light curve of EB V477\,Lyr (see Fig.\,\ref{477lyr}) is determined by eight parameters: $A_{1,2,3,4,5},D,C,\mathit{\Gamma}$, $n_{\rm e}=2$,\,$n_{\mathrm p}=3,\,n_{\mathrm c}=0$.

\subsection{The model of multicolour light curves}\label{multik}

The parameters of the above defined model functions, especially the amplitudes $A_k$, and parameters $C_{1,2}$, $D_{1,2}$ and  $\mathit{\Gamma_{1,2}}$ are generally functions of the wavelength $\lambda$.

In principle, we can use the one-colour models formulated in Sect.\,\ref{monmodel} separately, assuming that all of the parameters are wavelength dependent. Fortunately, it follows from our experience with modelling of LCs of hundreds of real systems and their physical models that it is not necessary to take response curves of different photometric passbands into account; we manage with their effective wavelengths only. It enables the association of photometric colours with different transparency widths, equal or close effective wavelengths (typically we are allowed to combine measurements done in $V$ and $y$).
\begin{figure}[ht]
\centering \resizebox{0.90\hsize}{!}{\includegraphics{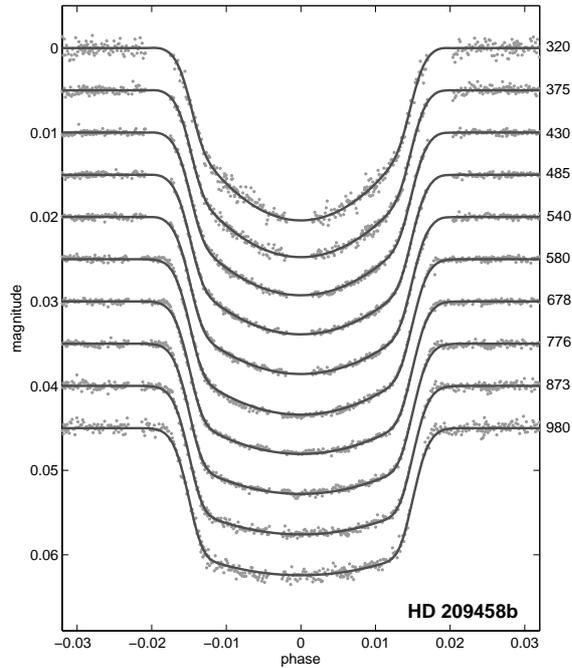}}
\begin{center}
\caption{\small Ten-colour transit LCs for HD\,209458b taken from \citet{knut}. Effective wavelengths of individual
colours are in nm. For the description of all 10 light curves we need only 9 parameters  \citep[courtessy of][]{mikcon}}  \label{knut}
\end{center}
\end{figure}

\begin{figure}
\centering \resizebox{0.83\hsize}{!}{\includegraphics{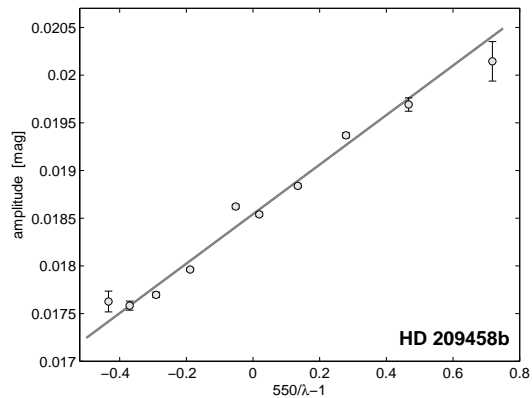}}
\caption{\small{The dependence of the amplitude $A$ of transit of HD\,209458b (see Fig.\,\ref{knut}) on $\mathit{\Lambda}=550/\lef-1$ proves the validity of the approximation Eq.\,(\ref{multi}) for that case.}} \label{ampl}
\end{figure}
On top of that the dependencies of model parameters on the effective wavelength $\lef$ are typically smooth, mostly monotonic, so we can approximate them by low-order polynomials of the dimensionless parameter $\mathit {\Lambda}$; $\mathit{\Lambda}=\lambda_0/\lambda_{\mathrm {eff}}-1$, where $\lambda_0$ is an arbitrarily selected central wavelength of
the data set \citep[][see also Fig.\,\ref{ampl}]{mikcon}:
\begin{align}\label{multi}
&\textstyle A_{k}=\sum_{j=1}^{g_{ {A\,k}}}a_{kj}\,\mathit{\Lambda}^{j-1},\quad
C_l=\sum_{j=1}^{g_{C\,l}}c_{lj}\,\mathit{\Lambda}^{j-1},\\
&\textstyle D_l=\sum_{j=1}^{g_{D\,l}}d_{lj}\,\mathit{\Lambda}^{j-1},\quad
\mathit{\Gamma}_{l}=\sum_{j=1}^{g_{\mathit{\Gamma} l}}\gamma_{lj}\,\mathit{\Lambda}^{j-1}, \nonumber
\end{align}
where $g_{Ak}, g_{Cl}, g_{Dl}, g_{\mathit{\Gamma} l}$, $l=1,2$ or $l=1$,
are the numbers of degrees of freedom of the corresponding
parameters of the model. The standard set of the one-colour
LC model parameters of EBs (see Sect.\,\ref{monmodel}):
$\{A_k,C_l,D_l,\mathit{\Gamma}_l\}$ can be considered as the special
case of the multicolour decomposition Eq.\,(\ref{multi}) for $\lambda_0=\lef, \mathit{\Lambda}=0$, $j=1$:
$\{a_{k1},c_{l1},d_{l1},\mathit{\gamma}_{l1}\}$.

The set of relations in Eq.\,(\ref{multi}) enables the calculations of all parameters $\{A_k(\mathit{\Lambda}),C_l(\mathit{\Lambda}),D_l(\mathit{\Lambda}), \mathit{\Gamma}\!_l(\mathit{\Lambda})\}$ needed for calculation of the model of a LC in any photometric band characterized by the parameter $\mathit{\Lambda}$.

Fig.\,\ref{knut} shows the fit of the extreme 10-colour photometry (350-980 nm) of an exoplanet transit \citep{knut} -- here we need 9 parameters, namely $A=a_1+a_2\mathit{\Lambda}$, $D=d_1+d_2\mathit{\Lambda}$, $C=c_1+c_2\mathit{\Lambda}+c_3\mathit{\Lambda}^2$, $\mathit{\Gamma}=\gamma_1+\gamma_2\mathit{\Lambda}$, where $\mathit{\Lambda}=550/\lambda_{\mathrm{eff}} -1$ \citep{mikcon}.

\begin{figure}
\centering \resizebox{0.92\hsize}{!}{\includegraphics{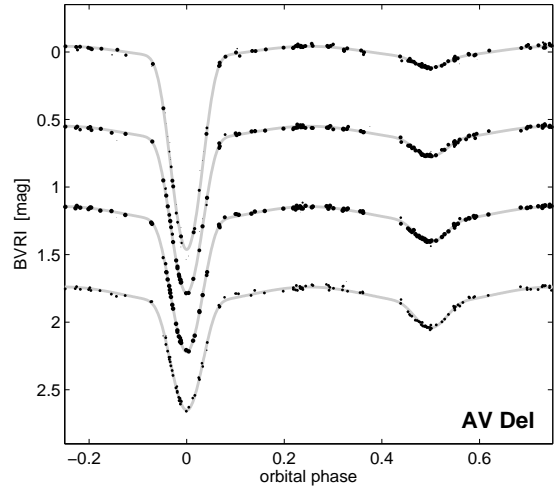}}
\caption{\small{The fit of the \textit{BV\!R}$_{\mathrm{c}}$\textit{I}$_{\mathrm{c}}$ LCs of a semidetached short periodic `cool Algol' AV Del. Dark dots are observations, grey lines is the 13-parameter phenomenological fit. For details see Table\,\ref{prm}} and Sect.\,\ref{select}.} \label{avdel}
\end{figure}

The fit of \textit{BV\!R}$_{\mathrm{c}}$\textit{I}$_{\mathrm{c}}$ proper LCs of AV Del (see Fig.\,\ref{avdel}) needs only nine parameters, namely $D_1=D_2=D$, $\mathit{\Gamma}_1=\mathit{\Gamma}_2=\mathit{\Gamma}$, $C_1=C_2=C$, $a_{11},\,a_{12},\, a_{21},\,a_{22},\,a_{41},$ and $a_{42}$, $n_{\mathrm e}$=2, $n_{\mathrm p}=2, n_{\mathrm c}=0$, $A_3 \equiv 0$. See also Table\,\ref{prm}.

\subsection{Brief description of selected eclipsing systems}\label{select}

TrES-3b is an extrasolar planet orbiting the star GSC\, 03089-00929 with a period of 31 hours. It belongs to the hot Jupiters which are undergoing orbital decay due to tidal effects. For the LC inspection (Fig.\,\ref{tres3}) $R$ photometry of 15 transits containing 2820 individual measurements in total \citep [courtesy][]{vanko} was used. The parameters of the LC fit of TrES-3b are in the legend of Fig.\,\ref{tres3}.

HD\,209458 was the first star found to have a transiting planet \citep{charb,henry} and remains the second brightest star known to have a transiting
planetary companion. \citet{knut} obtained 1066 spectra over four distinct transits with the STIS spectrometer on HST allowing to synthesize LCs in 10 spectrophotometric bandpasses in 290--1030 nm \citep[see also Sect.\,\ref{multik}, Fig.\,\ref{knut} and][]{mikcon}.

Table\,\ref{prm} contains the parameters of the phenomenological fit of LCs and some other information on the following selected eclipsing binaries:
\begin{itemize}
\item AR Aurigae, a prototype of a detached EB, \citep[][]{con}.

\item EK Comae, an overcontact, spotted EB with a short orbital period, \citep[][]{samec}.

\item AV Delphinis, a `cool Algol' consisting of a F type primary on the main sequence and a K subgiant filling its Roche lobe \citep[][]{mader}.

\item V477 Lyrae, an unusual, detached EB consisting of a very hot and luminous nucleus of the planetary nebula as a primary component and a solar type star as secondary \citep[][]{pol}.
\end{itemize}
\noindent Table\,\ref{prm} shows that the fit of LCs of the above mentioned stars by our `hi-fi' models, quantified by the ratio $\rho$ is nearly the same or better than in the case of the fit of BM3 (CALEB) physical model \citep{BM3}.

\section{Phenomenological model solution}\label{solution}

\subsection{Finding of model parameters and their uncertainties}\label{LSM}

The procedure for finding model parameters is based on the simultaneous mathematical processing of all relevant photometric data consisting of individual photometric observations, including barycentric julian date of the $i$-th measurement $t_i$, the measured magnitude or magnitude difference corrected for possible trend(s) during nights or seasons $y_i$, and the estimate of its uncertainty $\sigma_i$. Furthermore, we should know the effective wavelengths of the photometric filter used, $\lambda_{\mathrm{eff}i}\ \mathrm{or}\ \mathit{\Lambda}_i$, and submission of an individual observation to one of the observational subsets $r_i$ (see Sect.\ref{instrum}).

For simplicity we shall assume that the shapes of LCs are constant and the variability of an object is described by the unique model function Eq.\,(\ref{general}), consisting of the instrumental term $Y_{r0}(t,\lef)$ (Eq.\,(\ref{instrument})) and the intrinsic phenomenological ES model LC function $F(\vartheta,\lef)$, specified in Sect.\,\ref{fenec}. The phase function $\vartheta(t,P_0, M_0, \dot{P}_0,\ \mathrm{or}\ \{\oc\},\ldots)$ is a function of time and some free parameters of a variety of period models offered in Sect.\,\ref{gen}. The result of the solution - the full set of $g$ free parameters of the complete model including the estimate of the parameter uncertainties was evaluated simultaneously using the non-linear least square method by minimising the quantity $\chi^2$ by the standard technique (using tried Newton-Raphson method of non-linear equation solution) well described in e.g. \citet{press,hayashi,hart,mikcu}. With a good initial estimate of the parameter vectors the iterations converge fairly quickly.

\begin{figure}
\centering \resizebox{1.1\hsize}{!}{\includegraphics{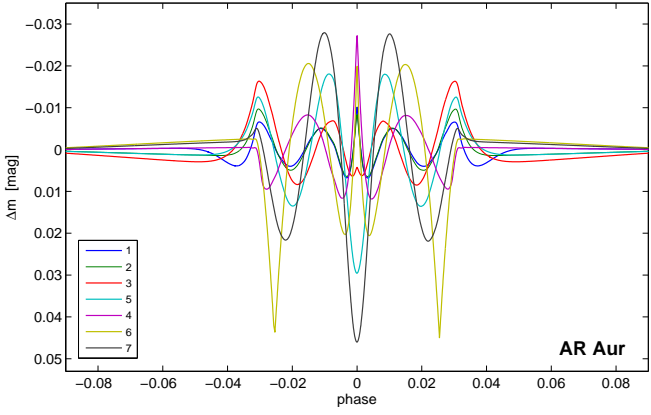}}
\caption{\small{The difference in mag between the physical model primary minimum of AR\,Aur and various alternatives of phenomenological basic function listed in Table\,\ref{srov}. }} \label{arsrov}
\end{figure}

All estimates of uncertainties of model parameters were computed using formulae taking into account that our models fit phase curves of EBs with uneven accuracy. Since the models are not orthogonal, uncertainties of the functions of model coefficients (typically the fits of LCs or minima times) should be computed by the general law of uncertainty propagation assuming also correlations among individual coefficients \citep[see e.g.][]{bevington,mikzej}. It is advisable to orthogonalize the models at least in the ephemeris parameters, in accordance with what we did in \citep[][]{mikort,mik901}.

\subsection{The selection of an optimal model of light curves }

\begin{table}[ht]
\caption{\small{The list of alternatives for several phenomenological model functions of eclipses.}}\label{srov}
\begin{center}\tiny
\begin{tabular}{lccl}
\hline \hline \rule{0pt}{1.07em}
&$F_{\mathrm{e}}(\varphi)$ & $s$ &parameters\\
&&mmag&\\
 \hline
1&$A\hzav{1\!+\!C\zav{\frac{\varphi}{D}}^2}\left\{1\!-\!\left\{1\!-\! \exp\hzav{1\!-\!\cosh\zav{\frac{\varphi}{D}}}\right\}^ {\mathit{\Gamma}}\right\}$&3.4 &$A,D,\mathit{\Gamma},C$\\
2&$A\left\{1\!-\!\left\{1\!-\!\exp\hzav{1\!-\!\cosh\zav{\frac{\varphi}{D}}}
\right\}^{\mathit{\Gamma}}\right\}$&3.9&$A,D,\mathit{\Gamma}$ \\
3&$A\left\{1\!-\!\left\{1\!-\!\exp\hzav{-\frac{1}{2}\zav{\frac{\varphi}{D}}^2}
\right\}^{\mathit{\Gamma}}\right\}$&5.5&$A,D,\mathit{\Gamma}$\\
4&$A\ \mathrm{Real}\zav{1\!-\!\left |\frac{\varphi}{D}\right |^\mathit{\Gamma}}^{3/2}$&6.9 &$A,D,\mathit{\Gamma}$\\
5&$A\exp\hzav{-\frac{1}{2}\zav{\frac{\varphi}{D}}^2}$&11&$A,D$ \\
6&$A\,\zav{\frac{1}{2}\!-\!\left|\frac{\varphi}{2\,D}\right|+\left|\left| \frac{\varphi}{2\,D}\right|\!-\!\frac{1}{2}\right|}$&16&$A,D$ \\
7&$A\exp\hzav{1\!-\!\cosh\zav{\frac{\varphi}{D}}}$&18&$A,D$ \\
\hline\\
\end{tabular}
\end{center}
\label{table}
\tiny{The fidelity of individual alternatives are illustrated by the Fig.\,\ref{arsrov} and quantified by the scatter $s$  of the fit of the `real' light curve of primary minimum of AR\,Aur, simulated by its physical model. The model function number 4 was suggested by \citet{and10,and12}, the function `Real' extracts a real part of the argument.}
\end{table}

The models of light curves should be tailored to the studied object/objects, available data, and the purpose of fitting the data. Especially, the number of used free parameters should be as few as possible, however without any serious influence on the accuracy and reliability of the results. It follows from our experience that it pays off to adhere to the following general principles:
\begin{enumerate}
\item It is always advantageous to process all available data simultaneously and not divided into parts. It is also valid for the determination of mid-eclipse times of individual eclipses where we should use the Eq.\,(\ref{oc}).
\item We have to pay attention to the right weighting of entered data as it is required by the used $\chi^2$ regression. If our original data do not content individual uncertainties, we have to estimate them iteratively from the scatter of residuals $\{\Delta y_i\}$ for appropriately created data subsets.
\item Fixing of LC model parameters, if they are known from previous analyses, is also advised. In addition, neglecting of all model terms whose amplitudes are less or comparable with their uncertainties, is recommended.
\item The number of free parameters can also be reduced by using simpler model functions of eclipses listed in Table\,\ref{srov}. Suggested model functions are compared in Fig.\,\ref{arsrov}.
\item The complexity of the selected phenomenological model should correspond to the purpose of the modelling. Several tasks (e.g. the basic classification of LCs) allow for the use of simple, unified models with basic parametric outfit.
\end{enumerate}
Unfortunately, the effort of using the optimum `hi-fi' phenomenological models considerably encumbers automation of the computational process. The diversity of real light curves of particular ESs requires that they be solved individually.

\section{Period analysis of XY Bootis in 1955--2009 } \label{xyperchng}

The above described phenomenological modelling of periodic variable stars has been developed first and foremost for the sake of the simultaneous period analysis of all available data containing phase information. Such data are in the case of ESs standardly times of light minima, LCs or their segments and radial velocity curves. The majority of studies of ES period changes were based on the analysis of \oc\ diagrams constructed using the times of light minima, where these timings were determined individually directly by observers not taking into account LCs of the star observed before. Several other studies were based only on the analysis of LCs obtained during several years. Measurements of radial velocities were used, namely, for the solution of the geometry of the system.

The simultaneous analysis of data of various kind is possible using an extended version of the contemporary sophisticated codes for the physical solution of eclipsing systems \citep[see e.g.][]{had,van,wilson14}. The concept of EB period analyses without using \oc\ diagrams and the physical solution of the system was outlined (i.a.) in \citet{miklijiang,miktatry, mikeas}, and studies of period changes of individual EBs \citep{zhudd,zhubs,mikbs}. The results of the preliminary report on the ephemeris of AR\,Aur -- a star with the light time effect \citep{mikar} -- were referred to and extensively discussed by \citet{wilson14}, who used AR\,Aur as an example of systems with a third companion.

\citet{wilson14} also studied the W UMa type eclipsing binary XY\,Bootis as a prototype of a close double star with a nearly steady period change. The study clearly proves the advantage of simultaneous processing of all relevant data. That is why we want to compare the results of this sophisticated study describing period change of XY\,Boo during the time interval 1955--2009 with our results based on strictly the same observational data.

\subsection{XY Bootis}
Eclipsing binary XY Bootis (= BD$+20^{\circ}2874$ = HIP 67431; $V_{\mathrm{max}}=10.3$\,mag; Sp. F5V) was discovered as a variable star by \citet{hoff}. \citet{ces50,ces54} observed the star visually and classified it as an eclipsing binary of W UMa-type.

The first photoelectric observations were obtained by \citet{hind}. \citet{wood} reanalysed Hinderer's data and found the true period of the star $P$\,= $0\fd37054$. \citet{bin} observed six minima of \hvezda\ in $BV$, improved the ephemeris $(M_0=2440389.7321,\ P=0\fd37055)$, and revealed remarkable increase in the period. \citet{winkler} observed \hvezda\ also in 1976 in $BV$ and derived three other times of minima (see Table\,\ref{xytmin}). \citet{awa} gave two new times of light minima and confirmed period increase. The history of the \hvezda\ investigation from its discovery to 1998 is described more extensively in \citet{mol}. \citet{lean} measured spectroscopically the radial velocity of both components and estimated the mass ratio $q=0.16\pm0.04$.

\begin{figure}
\centering \resizebox{0.90\hsize}{!}{\includegraphics{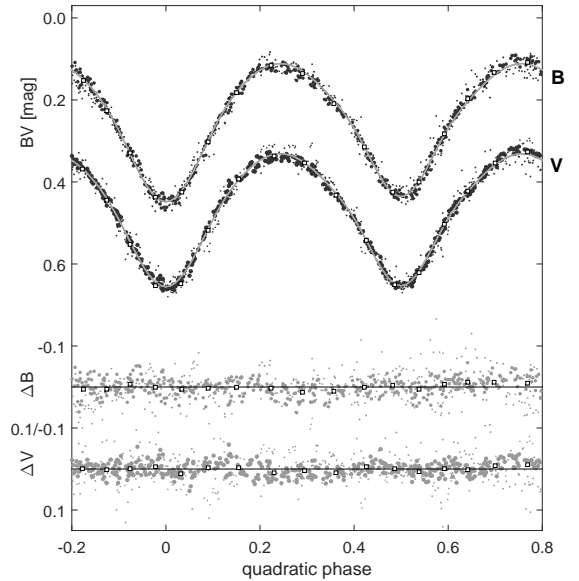}}
\caption{\small{$BV$ curves of \hvezda\ (see the list in Table\,\ref{xyobs}). The phase is plotted according to our quadratic ephemeris (see Table\,\ref{xypar}). The areas of points are proportional to their weights. Open squares are normal points (each of them represents the mean of about 50 measurements). Grey lines are fits by our phenomenological model (see Eq.\,(\ref{xyLCeq}) and Table\,\ref{xycrv}). $\Delta B$ and $\Delta V$ display residuals of $BV$ magnitudes and normal points from the LC phenomenological model. The scale of residuals is two times larger than the measure of \textit{BV} light curves.}} \label{xyLCfig}
\end{figure}

The first detailed period study was done by \citet{mol} who used 43 moments of minimum spanning the interval 1944--98 and calculated the quadratic ephemeris. The found record-breaking period increase $\dot{P}$\,=\,1.67(5)$\times 10^{-9}$=\,5.3\,s per century was explained by mass transfer of $1.34\times 10^{-7}$\,M$_{\odot}$\,yr$^{-1}$ from the secondary to the primary component. The results were confirmed and improved by \citet{yang} who found $\dot{P}$\,=\,1.711(6)$\times 10^{-9}$=~5.4~s per century on the basis of standard \oc\ analysis of 54 minima times.

\citet{wilson14} combined in their calculations of the quadratic ephemeris of \hvezda\ phase information hidden in \textit{BV} LCs obtained by \citet{bin,winkler,awa}, $RV$ curves of \citet{lean}, all times of minima listed in \citet{yang} and other 33 minima timings collected in Table 1 of their article covering the interval 2005--09. Using all of those data they found the mean period increase $\dot{P}$ = 1.6348(8)$\times 10^{-9}$ = 5.159(24) s\,per century. They also noticed some oscillation from JD~2\,448\,000 to 2\,455\,000 with no indication of periodicity. We have now collected many other observations proving the complexity of the \hvezda\ period variations. Nevertheless, for the sake of the comparison of the effectiveness of our method, we used in the following small study exclusively those data used by \citet{wilson14}.

\subsection{Phenomenological model of \hvezda\ variability}\label{xymod}
We assumed, similarly as \citet{wilson14}, that the instantaneous period $P(t)$ of \hvezda\ is lengthening with the constant rate $\dot{P}(t)=\mathrm{const.}$ Then the phase function $\vartheta(t)$ and the prediction of primary minimum times $\mathit{\Theta}(E)$ can be approximated (according to Eqs.\,(\ref{thetalaurin}), (\ref{Thetalaurin})) by simple relations:
\begin{equation}
\vartheta=\vartheta_1-\dot{P}\,\frac{\vartheta_1^2}{2};\quad \label{xytheta}
\mathit{\Theta}_{\mathrm{min}}=M_0+P_0\,E+\textstyle{\frac{1}{2}}P_0\,\dot{P}\,E^2,
\end{equation}
where $\vartheta_1=(t-M_0)/P_0$, $M_0$ is the JD timing of the basic primary minimum - the origin of counting of epochs $E$, $P_0=P(t=M_0)$ is the instantaneous period at the time $t=M_0$. Integer doubling of epoch E  is done (even for times of primary minima, odd for times of secondary minima)\footnote{\w\ used for modelling of phase function $\vartheta$ and its inversion $\mathit{\Theta}$ an exact solution of the basic equation, Eq.\,(\ref{phasefundef}): $\vartheta(t)\!=\!\dot{P}^{-1}\ln \hzav{ 1+\dot{P}/P_0\,(t-M_0)}$, $\mathit{\Theta}(E)\!=\!P_0/\dot{P} \hzav{\exp(E \dot{P}-1)-1}+M_0$. Nevertheless, our models for $\vartheta$ and $\mathit{\Theta}$ (see Eqs.\,\ref{xytheta}) do not differ by more than $1.9\times10^{-5}$ in the phase function and 0.3 s in time prediction from the exact ones. So they can be considered to be identical.}.

Light curves of \hvezda\ with almost equally deep minima (see Fig.\,\ref{xyLCfig}) agree with W UMa classification. We found that all of them can be well fitted by the following phenomenological model with nine free parameters: $a_{11},a_{12},a_{21},a_{22},A_3,A_4,D_1,D_2$, and $\mathit{\Gamma}$,
\begin{eqnarray}
&\displaystyle F_{\mathrm{LC}}(\vartheta)=\sum_{k=1}^2 (a_{k1}+a_{k2}\mathit{\Lambda})\left\{1\!-\!\left\{1\!-\!\exp\hzav{1\!- \!\cosh\zav{\frac{\varphi_k}{D_k}}}
\right\}^{\mathit{\Gamma}}\right\}+\nonumber \\
&+\,A_3\cos(4\,\pi\,\vartheta)+A_4\cos(6\,\pi\,\vartheta);\label{xyLCeq}\\
&\displaystyle  \varphi_k=\hzav{\vartheta-(k-1)/2} -
\mathrm{round}\hzav{\vartheta-(k-1)/2},\quad \mathit{\Lambda}=\frac{550}{\lambda_{\mathrm{eff}}}-1. \nonumber
\end{eqnarray}
The results achieved using this phenomenological model of \hvezda\ $BV$ light curves are denoted as `Model\,I'.

\begin{figure}
\centering \resizebox{0.80\hsize}{!}{\includegraphics{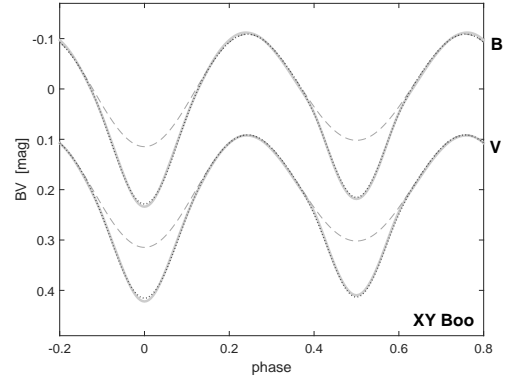}}
\caption{\small{Models of $B$ and $V$ LCs of \hvezda. Grey lines are \textit{BV} model light curves simulated by the \citet{wilson14} physical model, dotted lines correspond to our phenomenological model (see Eq.\,(\ref{xyLCeq}) and Table\,\ref{xypar}); dashed lines are hypothetical \textit{BV} LCs of XY\,Boo without mutual eclipses.}} \label{xyLCmodfig}
\end{figure}

Other LC templates were $BV$ synthetic LCs computed by the freely accessible W-D 2013 code with physical parameters of \hvezda\ published in \w. The results obtained using these $BV$ LC templates are denoted as `Model\,II'. Fig.\,\ref{xyLCmodfig} displays that the difference between the found phenomenological and synthetic $BV$ model LCs are only marginal; numerically it is also documented in Table\,\ref{xycrv}.

\begin{table}\tiny
\begin{center}
\caption{\small{Nine parameters of phenomenological (this paper) and synthetic LCs \citep{wilson14} models.}}
\begin{tabular}{lll|lll}
\hline
par.& Model I& W\&VH & prm & solution I & W\&VH \\
\hline
$a_{11}$ [mag]& 0.102(4)& 0.098 &$A_{4}$ [mag]& 0.0064(6) & 0.0075 \\
$a_{12}$ [mag]& 0.112(4)& 0.104 &$D_1$ &0.0627(21) & 0.059\\
$a_{21}$ [mag]& 0.050(10)& 0.056 & $D_2$ &0.0558(20) & 0.053\\
$a_{22}$ [mag]&0.007(10)& 0.048 & $\mathit{\Gamma}$ &0.97(6) & 0.95 \\
$A_{3}$ [mag]& 0.1081(19)& 0.111&         &    & \\
\hline \label{xycrv}
\end{tabular}
\end{center}
For the meaning of parameters see Eq.\,(\ref{xyLCeq}).
\end{table}

For the description of shapes of radial velocity $(RV)$ phase curves we used only a simple sinusoidal model (see Eq.\,(\ref{xyrveq})) neglecting the Rossiter-McLaughlin effect \citep{laugh}
\begin{equation}
RV_k(\vartheta)=V_{\gamma}+A_5 \zav{k-1-q}\,\sin(2\,\pi\,\vartheta),\label{xyrveq}
\end{equation}
where $k=1$ refers to the first component, whilst $k=2$ to the secondary one, $V_{\gamma}$ is so-called $\gamma$ velocity, $A_5$ is the amplitude of the difference in the radial velocity between the components, all in km\,s$^{-1}$, and $q=m_2/m_1$ is the ratio of component masses.

This approximation is fully sufficient for $RV$ observations of \citet{lean} that cover only about 15\,\% of the phase curve. Results are in the bottom part of Table\,\ref{xypar}.

\begin{table}\tiny
\begin{center}
\caption{\small Comparison of common parameters derived by \citet{wilson14} and us.}
\begin{tabular}{lccc}
\hline
  par. & W\,\&\,VH & Model I & Model II\\
   \hline
all\\
\hline
$\dot{P}$ &$1.6348(77) \times 10^{-9}$& $1.6328(52)\times 10^{-9}$& $1.6442(47)\times 10^{-9}$\\
$P_0$& $0\fd370\,560\,205(18)$& $0\fd370\,560\,213(14)$& $0\fd370\,560\,198(12)$\\
$M_0^*$  & 0.35163(28) & 0.35185(26) & 0.35143(27) \\
\hline
LC \\
\hline
$\dot{P}$ & $1.497(37) \times 10^{-9}$ & $1.506(24)\times 10^{-9}$& $1.502(25)\times 10^{-9}$ \\
$P_0$& $0\fd370\,559\,889(92)$& $0\fd370\,559\,913(70)$& $0\fd370\,559\,898(71)$\\
$M_0^*$  & 0.35162(29) & 0.35174(30) & 0.35164(30) \\
\hline
$T_{\mathrm m}$ & W\,\&\,VH & this paper \\
\hline
$\dot{P}$ & $1.729(27) \times 10^{-9}$ & $1.745(26)\times 10^{-9}$\\
$P_0$& $0\fd370\,559\,58(11)$& $0\fd370\,559\,86(10)$\\
$M_0^*$ & 0.3565(21) & 0.3490(11) \\
\hline
$RV$&W\,\&\,VH&this paper \\

\hline
$V_{\gamma}$ &$7.7\pm6.6$&$8\pm7$\\
$q$&$0.159\pm0.048$&$0.154\pm0.037$\\
$A_5$ &$286\pm24$& $282\pm23$\\
\hline
\end{tabular}\label{xypar}
\end{center}
$M_0^*= \mathrm{HJD}-2\,444\,716$, $V_{\gamma}$ and $A_5$ are expressed in km s$^{-1}$.
\end{table}

\subsection{Used data and their weights. Models I and II}

Our period analyses of \hvezda\ was based on 1770 individual data points (see the list in Table\,\ref{xyobs}) acquired by three different techniques divided into 9 groups with various scatters $\sigma_{\mathrm {I}}$ and $\sigma_{\mathrm {II}}$ with respect to phenomenological (I), and synthetic (II), model phase curves.

\begin{figure}
\centering \resizebox{0.86\hsize}{!}{\includegraphics{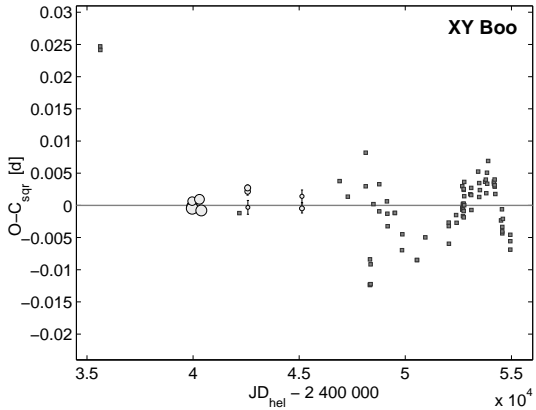}}
\caption{\small{\hvezda\ timing residuals from the quadratic ephemeris (with a $\dot{P}$ term). Published minima timings are denoted as squares, timings derived from LCs are circles which areas are proportional to their weight.}}\label{xyOCkva}
\end{figure}

Scatter $\sigma_{\mathrm {I}}$ used to be the same or a bit smaller than the scatter $\sigma_{\mathrm {II}}$ with respect to computed synthetic phase curves \citep{wilson14}. Scatters of radial velocity measurements -- $\sigma_{RV \mathrm{I}}$ and $\sigma_{RV \mathrm{II}}$ 28 and 30\,km\,s$^{-1}$ -- seems to be a good estimate of their values. The mean uncertainty of eclipse times\footnote{We did not distinguish between primary and secondary light minima because their depths are nearly the same.} $\sigma_{\mathrm{Tmin}}$ with respect to the model Eq.\,(\ref{xytheta}) is $0\fd0043=6.2$\,min. It should be mentioned that the inner uncertainty of individual times of minima is typically one order better. The uncertainty $\sigma_{\mathrm{Tmin}}$ is caused mainly by erratic fluctuations in the rate of the orbital period and the shape of LCs. The character of these changes is well displayed in Fig.\,\ref{xyOCkva}.

Scatter $\sigma_{\mathrm {I,\ II}}$ for an individual subset of data (see Table\,\ref{xyobs}) served as the basis for weighting of each used measurement and interconnection of data of various nature. It enabled applying of $\chi^2$ regression computation outlined in Sect.\,\ref{solution}. The computation represents the solution of 21 (case I) or 12 (case II) non-linear equations of 21 or 12 free parameters. The difference in the number of parameters needed is a result of the fact that in case II we adopted template LCs from paper \citet{wilson14}. This means that we fixed all 9 parameters that we needed for the determination of $BV$ template LCs.

\begin{table}\tiny
\begin{center}
\caption{\small{The list of used observational data including their source, number, specification, and scatter with respect to phenomenological model phase curves (Model I) and synthetic phase curve (Model II). }}
\begin{tabular}{lcrcll}
\hline
 source&type &N& spec&$\quad \sigma_{\mathrm I}$&$\quad \sigma_{\mathrm {II}}$\\
   \hline
 \citet{bin}&light&440&$B$&0.011 mag&0.011 mag\\
 &curves&441&$V$&0.0084 mag&0.0087 mag\\
  \hline
  \citet{winkler}&light&99&$B$&0.021 mag&0.020 mag\\
 &curves&109&$V$&0.022 mag&0.023 mag\\
\hline
  Awadala (1984)&light&280&$B$&0.021 mag&0.023 mag\\
 &curves&284&$V$&0.025 mag&0.026 mag\\
\hline
 McLean (1983)&radial&12&I&28 km\,s$^{-1}$&28 km\,s$^{-1}$\\
 &vel.&10&II&30 km\,s$^{-1}$&30 km\,s$^{-1}$\\
\hline
\citet{yang}&eclipse&75&I+II&0.0043 d&0.0043 d\\
+W\&V(2014)&timings\\
\hline
\end{tabular}\label{xyobs}
\end{center}
\end{table}

\begin{figure}
\centering \resizebox{1\hsize}{!}{\includegraphics{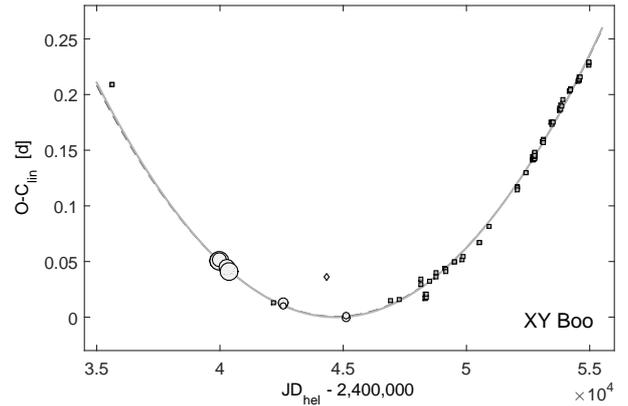}}
\caption{\small{\hvezda\ timing residuals from the linear ephemeris (without a $\dot{P}$ term). Published minima timings are denoted as squares, timings derived from LCs are circles which areas are proportional to their weight, and a timing derived from radial velocity curves is noted as a diamond.}} \label{xyOClin}
\end{figure}

In the first part of solutions I and II we obtained the parameters of the unified \hvezda\ ephemerides: $M_0$, $P_0$, and $\dot{P}$ (see Table\,\ref{xypar}), which can be compared with the same parameters listed in Table\,3 of \citet{wilson14} paper. Computed parameters of the $B$, $V$, and $RV$ phase curves according to these ephemerides are in Table\,\ref{xycrv} and Table\,\ref{xypar}, respectively. In the second part of the solutions we derived 12 virtual eclipse times and their uncertainties by means of the model Eq.\,(\ref{oc}) from $BV$ photometry and $RV$ data - see Table\,\ref{xytmin}, Fig.\,\ref{xyOCkva}, \ref{xyOClin}.

\subsection{Discussion of results. Comparison with W\&VH solution}\label{xyresults}

\subsubsection{General remarks}
We discuss here the results of three methods for period analysis that enables the processing of various type of data containing phase information. For their consistent comparison we used exactly the same observational data set specified in the paper of \citet{wilson14} and the assumption that the period $P$ rises uniformly with time: $P(t)=P_0+\dot{P}\,(t-M_0)$, where $\dot{P}=\mathrm{const}$.

Model I uses the phenomenological model for the $BV$ LCs (Eq.\,(\ref{xyLCeq})), whilst hybrid Model II uses synthetic LCs computed by \w. This technique is the one used e.g. by \citet{zasche14,zasche15}. Both  models assume strictly circular orbits of components and synchronous rotation. They also suppose that the shape of LCs given, namely, by the variable geometry of the orbiting system, are more or less constant.

\begin{figure}
\centering \resizebox{0.84\hsize}{!}{\includegraphics{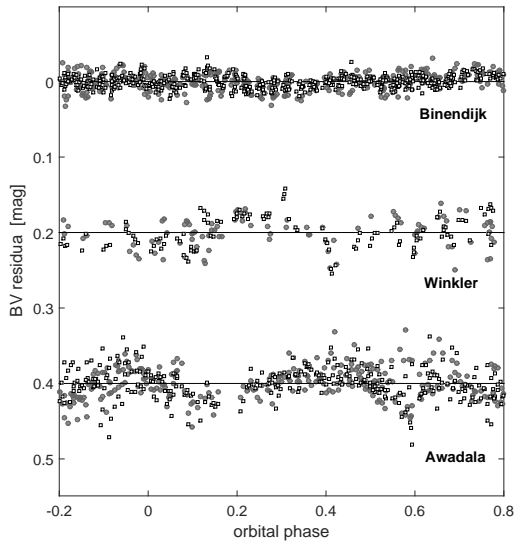}}
\caption{ \small{Residua of $BV$ light curves of \citet{bin,winkler,awa} clearly show variations in LC shapes. $B$ and $V$ residuals are denoted as circles and open squares, respectively.}} \label{xyLCresfig}
\end{figure}

On the contrary, detailed inspection of the LCs proves that there are apparent changes on various time scales \citep[see Fig.\,\ref{xyLCresfig} and $BV$ light curves in Fig.\,3 in][]{wilson14}.
Some of them can be attributed to instrumental effects like red noise \citep{pont} or incomplete detrending of LCs. Others are due to intrinsic changes of the object itself, such a chromospheric and spot activity or shifts due to unsteady mass transfer between components (see the right part of Fig.\,\ref{xyOCkva}).

\begin{table}\tiny
\begin{center}
\caption{\small Comparison of the eclipse timings of \hvezda.}\label{xytmin}
\begin{tabular}{lllrrc}
\hline
  publ.\,timing &Model\,I &Model\,II & (\oc)$_{\mathrm{I}}$[d] & epoch\   & aut\\
   \hline
 39\,950.8121&.81207(16)&.81207(16)&-0.00024&-12860.5&B\\
 39\,951.9243&.92427(14)&.92430(13)&0.00030&-12857.5&B\\
 39\,953.7763&.77635(13)&.77628(13)&-0.00038&-12852.5&B\\
 39\,953.9626&.96259(20)&.96261(21)&0.00058&-12852.0&B\\
 40\,298.9470&.94746(16)&.94745(16)&0.00089&-11921.0&B\\
 40\,389.7319&.73142(13)&.73145(13)&-0.00065&-11676.0&B\\
 42\,569.7091&.70875(55)&.70875(55)&0.00212&-5793.0&W\\
 42\,577.6745&.67530(65)&.67535(65)&0.00173&-5771.5&W\\
 42\,582.6769&.6750(12)&.6750(12)&-0.00108&-5758.0&W\\
 45\,131.3793&.3800(11)&.3800(10)&0.00047&1120.0&A\\
 45\,132.3056&.30470(80)&.30450(85)&-0.00126&1122.5&A\\
\hline
 44\,325.& .446(16)  & .446(16)&  0.03500   & -1055.0   & RV \\
  \hline
\end{tabular}
\end{center}
Published timings of \citet{bin}, \citet{winkler}, and \citet{awa} are compared with our results (only the fraction is given) and their uncertainties derived from LCs and RV curves. (\oc)$_{\mathrm{I}}$ is the residual of the found eclipse timing to quadratic ephemeris according to the Model II, $N$ is the number of photoelectric observations used for the timing determination.
\end{table}

\subsubsection{Unified ephemerides. Virtual eclipse times}\label{xyuniefem}

All quantitative results achieved by the three discussed models which can be directly compared are listed in Table\,\ref{xypar}. It is apparent that especially the increase of the orbital period $\dot{P}$, the period $P_0=P(t=M_0)$, and the time of the basic primary minimum $M_0$ derived from all data, from LCs only and from eclipse times, are identical within their uncertainties. The results derived from the analysis of radial velocity measurements are in very good concordance as well. We are convinced that this conclusion is valid not only for \hvezda\, but also for other eclipsing systems.

There is an apparent identical discrepancy found by this paper and \citet{wilson14} in the value of the mean deceleration $\dot{P}$ derived only from LC shifts and/or eclipse times ($T_{\mathrm m}$) (see Table\,\ref{xypar}, rows LC and
$T_{\mathrm m}$): $\dot{P}_{\mathrm {LC}}=1.506(24)\times10^{-9}$, $\dot{P}_{T_{\mathrm m}}=1.745(26)\times10^{-9}.$

The first possible explanation, that it is the result of different distribution of LCs and times data along the \oc\ diagram (see Fig.\,\ref{xyOClin}, \ref{xyOCkva}) and the inconstancy of $\dot{P}$ where $\ddot{P}>0$), has proven to be false. We tested the period models with $\ddot{P}$ term and found that $\ddot{P}=0(4) \times 10^{-15}$\,d$^{-1}$. The effect is then very likely caused by the presence of the pair of eclipse minima of \citet{hind} at the very beginning of the \oc\ diagram (see Fig.\,\ref{xyOCkva}) that significantly deviates from its parabolic prediction. However, our thorough inspection of the original Hinderer's measurements showed that both eclipse times are correct.

Models I and II enabled calculation of `virtual eclipse times' for selected subsets of photometric and radial velocity data using the relation Eq.\,(\ref{oc}). These times, including their uncertainties, are listed in Table\,\ref{xytmin} together with eclipse times\footnote{All the eclipse times were given without quoting their uncertainties.} published by authors of $BV$ photometries -- \citet{bin,winkler,awa}  The results as well as their uncertainties obtained by Model\,I and Model\,II are nearly identical. The last ones are very important as they enable the evaluation of real inner accuracy of individual times, the construction of precise \oc\ diagrams \citep{miklijiang}, and the discussion of subtle problems of stability of the period, variability of LCs etc. There is no systematic difference between the published eclipse timings (see the first column of Table\,\ref{xytmin}) and those derived by our Models.

The excellent agreement of comparable results obtained by all three models proves that they can be used as good alternatives in solving common tasks of the eclipse system period analysis.

\subsubsection{\textit{BV} and \textit{RV} phase curves}\label{xyphasecrv}

The differences between the phenomenological and the synthetic $BV$ light curves are insignificant -- they represent about 0.5\% of the amplitude of the light changes (see Fig.\,\ref{xyLCmodfig}). The parameters of the phenomenological model of both mentioned LCs are in Table\,\ref{xycrv}. It seems that phenomenological modelling is a good method for expressing common types of LCs using only a very small number of parameters.

The fit of the observed $BV$ light curves by Model I is fairly good (see Fig.\,\ref{xyLCfig}) -- scatter of normal points in $B$ and $V$ are better than 0.004 and 0.003 mag. The larger part of this scatter is caused by the erratic variability of LCs due to stellar activity and inconstant O'Connell effect. The changes of LC shapes are clearly visible in the phase diagram of $BV$ residuals (Fig.\,\ref{xyLCresfig}) for all three photometric data sets.

The detailed inspection of measurements of \citet{bin} shows weak and variable O'Connell effect (data from 1968 and 1969 displayed O'Connell effect of the opposite sign), whilst residuals of \citet{winkler} and \citet{awa} are dominated by double wave going in antiphase. It seems that such seasonal variations in light curve shapes are quite common. Sometimes they can be much more dramatic \citep[see e.g. changes in LCs of notorious XY\,UMa,][]{lister}. These LC changes lowered the accuracy of the fit and eclipse timing determination.

Yet unpublished observations of \hvezda\ obtained during one month \citep{mikxy} prove that the time scale of LC changes is usually shorter than several days.

\section{Conclusions } \label{conclusions}

%\subsection{Experiences with phenomenological modelling of eclipsing systems}
%We are convinced that p

Phenomenological modelling presents an admissible alternative for the solution of selected ES research tasks. We state that:
\begin{enumerate}
\item The estimation of parameters and their uncertainties obtained by our phenomenological (and also hybrid) modelling and other well-proven methods, including solutions by sophisticated physical models of eclipsing systems, are almost the same (see e.g. Sect. \ref{select} and \ref{xyresults}).
\item Phenomenological modelling is based on the minimization of the $\chi^2$ sum, which enables the simultaneous processing of different sources of phase information (complete LCs and their segments, radial velocity curves, individual mid-eclipse times etc.).
\item Simple model function of eclipses (Eq.\,(\ref{modelecl})) can be also used for good determination of mid-eclipse times and their uncertainties of individual observations of stellar eclipses \citep[our model has been standardly used for the determination of time minima of original observations of EBs by Variable star and Exoplanet Section of Czech Astronomical Society since 2011, ][]{brat}
\item As a by-product of the phenomenological modelling we obtain the list of virtual minima times derived from LC data which help us i.a. to quickly check the selected phenomenological model of LCs or period variability (see Sect.\,\ref{xyuniefem}).
\item Light curve model parameters can be used for the apt description of both one-colour and multicolour LCs of eclipsing systems. For example, shapes of ten curves of HST spectrophotometry (320--980 nm) of the exoplanet transit of HD~209\,458b, taken from \citet{knut}, are determined by only nine parameters \citep[see][]{mikcon}. We can use this application e.g. for description and classification of observed LCs of eclipsing systems.
\item Knowing the template of LCs (from physical models or observed LC of superior quality) we can quickly modify the presented phenomenological model and establish a hybrid model (Model II in Sect.\,\ref{xyperchng}) with the diminished number of free parameters. As we showed in Sect.\,\ref{xyperchng}, the results of this approach are the same results as physical or pure phenomenological modelling. However, application of the latter method is much faster.
\item Phenomenological (and also hybrid) modelling of an ES could solve standard tasks of ES research (based on every sources of phase information), especially the improvement of ES ephemeris for standard period models of:
    \begin{itemize}
    \item systems with constant period,

    \item systems in a steady regime of mass and angular momentum transfer between components (e.g. XY Boo, Sect.\,\ref{xyperchng}),

    \item systems with erratic changes of period -- phase function is here approximated using so-called \oc\ time shifts of the observed phase curves versus the predicted LC derived by the period model with fixed parameters (typically $P_0,M_0$ or $P_0,M_0,\dot{P}$) - see Eq.\,(\ref{oc}), and Sect.\,\ref{xyuniefem},

    \item eclipsing systems influenced by the gravitational attraction of a third body \citep[see the preliminary paper about AR\,Aur in][and the discussion in Wilson \& VanHamme 2014]{mikar}\footnote{The detailed study about the triple star AR\,Aur is in preparation. For formulae see \citet{liska}.}.
    \end{itemize}

\end{enumerate}

\noindent Phenomenological modelling of eclipsing systems presented above has also its disadvantages and limitations:
\begin{itemize}
\item[$\bullet$] The method does not provide direct information about the geometry and physics of eclipsing systems, especially the inclination of the system, relative radii, temperatures and the form of components.
%    Nevertheless, the prepared statistical studies promise to find some correlations among the above mentioned and phenomenological parameters.
\item[$\bullet$] The phenomenological models of EBs light curves in their proposed form are not able to achieve an accuracy better than 0.5\,\% of their total amplitude of them (the accuracy of Kepler or CoRoT light curves is better by one or more orders). Nevertheless, such uncertainty is fully acceptable for many application.
\item[$\bullet$] It may be a bit troublesome for beginners to select the optimum model of LCs. Using an inappropriate model may lead to nonconvergent or bad solutions.
\item[$\bullet$] Phenomenological modelling of ESs has not been made available in such a user-friendly format as other modern physical models \citep[e.\,g.][]{had,prsa08,prsa11,wilson} until now.
\end{itemize}

\noindent  Phenomenological modelling, influenced by modern physical models, has been developed gradually since 2008 \citep{mik901}, its elements and principles were used in the majority of papers of this paper's author. The detailed applications of the presented bases of the method will be published elsewhere.

\begin{acknowledgements}
The author is very indebted to many of his collaborators, namely to M. Zejda, S. de Villiers, M. Chrastina, M. Jagelka, J. Krti\v{c}ka, J. Li\v{s}ka, E. Paunzen, T. Pribulla, S.-B. Qian, M. Va\v{n}ko, R. E. Wilson, L.-Y. Zhu, and others, for their help with this arduous theme. The work on the paper was partly sponsored by the grants Ministry of education of Czech Republic II LH14300 and GA\,\v{C}R 13-10589S.
\end{acknowledgements}

\end{document}